\documentstyle[12pt,epsf]{article}
\begin{document}
\setcounter{page}{1}
\pagestyle{plain}
\setcounter{equation}{0}
%
%
%
\ \\[12mm]
\begin{center}
    {\bf BOSE-EINSTEIN CONDENSATION IN \\[1mm]
 DISORDERED
EXCLUSION MODELS AND \\[1mm] RELATION TO TRAFFIC FLOW}
\\[15mm]
\end{center}
\begin{center}
\normalsize
        M.\ R.\ Evans\footnote{
Royal Society University Research Fellow}\\[13mm]
        {\it Department of Physics and Astronomy\\
        University of Edinburgh\\ Mayfield Road, Edinburgh EH9 3JZ, U.K.}
\end{center}
\vspace{1.0cm}
{\bf Abstract:} 
A disordered version of the
 one dimensional asymmetric exclusion model where
the particle hopping rates are quenched random variables is studied.
The steady state is  solved exactly
by use of a matrix product.
It is shown how
  the phenomenon of Bose
condensation whereby a finite fraction of the empty sites
are condensed in front of the slowest particle may occur.
Above a critical density of particles a phase transition occurs
out of the low density phase  (Bose condensate)
to a high density phase.
An exponent describing the decrease of the steady state velocity
as the density of particles goes above the critical value is calculated
analytically and shown to depend on the distribution of
hopping rates. The relation to traffic flow models is discussed.
\\[25mm]
2/6/1996 \\
Submitted to Europhys. Lett.
\\[3mm]
\rule{7cm}{0.2mm}
\begin{flushleft}
\parbox[t]{3.5cm}{\bf Key words:}
\parbox[t]{12.5cm}{ asymmetric exclusion process,
                   matrix product ground states,\\ disorder, Bose condensation}
\\[2mm]
\parbox[t]{3.5cm}{\bf PACS numbers:} 05.70Fh,  05.70Ln, 05.40+j
\end{flushleft}
\normalsize
\thispagestyle{empty}
\mbox{}
\pagestyle{plain}
%
%
%
\newpage
\baselineskip=18pt plus 3pt minus 2pt
\setcounter{page}{1}
\setcounter{equation}{0}
\noindent {\bf Introduction:}
Nonequilibrium systems may be defined as those evolving to
some dynamical rule which a priori does not
respect detailed balance with respect to an energy function.
The interest in such systems is that they describe a broader class
of phenomena,
for example  driven diffusive systems \cite{SZ} and growth processes \cite{HHZ},
 than can be handled  within the traditional framework
of  equilibrium statistical mechanics.
The  asymmetric exclusion model is a
prototypical nonequilibrium model.
In various guises it has been used to model
interface growth, polymer dynamics and
traffic flow \cite{SZ,HHZ,vLK,Traffic}. It comprises
classical particles stochastically making nearest neighbour
hops with hard core
exclusion and a preferred direction imposed. Through iteration of the dynamical rules
a nonequilibrium steady state is attained. Even in one dimension
steady states of asymmetric exclusion processes have
 revealed such diverse  phenomena as
long range correlations, shock formation,  boundary-induced
phase transitions and symmetry breaking \cite{Krug,DEHP,SD,sclass,EFGM}.
Significant progress has been made in understanding the one dimensional
model through the exact solution of  steady states via a matrix technique
\cite{DEHP}.
Recently  this technique has been extended from
 continuous time (random sequential) dynamics  to
a parallel  updating scheme \cite{Haye}.

In the present work the effects of disorder in the model are considered.
Specifically, each particle has its own intrinsic rates for hopping
forwards and backwards which  are taken as random variables.
Previously, single inhomogeneities
such as a fixed blockage \cite{Blockage}
or a moving blockage \cite{BerChine} were studied
and phase transitions found.
In another context a repton model with two allowed hopping rates 
was solved  and a mapping to a Bose gas noted  \cite{vLK}.
 Here we employ a similar mapping and
report that in the fully disordered model
 studied, Bose-Einstein condensation  occurs.
In the limit of a single inhomogeneity the condensation transition
corresponds to a limit
of  \cite{BerChine}.

 To  understand, at a descriptive level, the Bose condensation found here 
it is useful
to visualise a caricature of traffic flow. Consider
a one lane road with  cars moving along, each at its own
chosen speed, and no overtaking allowed. After some time a tailback will
be formed behind the slowest car with all the other cars' speeds reduced
to that of the slowest. However, if the density of the cars on the road is increased, for example, by cars merging with the road, a congested situation may
be reached wherein the speed of all cars, regardless of their
 own preferred speeds, is limited by the high density
of traffic.
To see the analogy with Bose condensation one relates  the
distribution of available road between cars
to the distribution of
Bosons within the states of an ideal Bose gas.
In the low density case a finite fraction of the available open road
is condensed in front of the slowest car whereas in the high density
situation the available road is spread thinly between all cars. 

In the following the analogy between a phase transition in the model
considered and Bose condensation will be made exact through
the exact calculation of the steady state of the model. This is performed
with the aid  of a matrix product ansatz.
A critical exponent describing how the velocity of  a particle 
decreases as the density goes above the critical
value is calculated exactly and related
to the distribution of car
velocities.

\noindent {\bf Model Definition and Construction of Steady State}:
The model studied here is defined as follows. 
Consider $M$ particles labelled $\mu =1,\ldots, M$ hopping
on a lattice of size $N$ with periodic
boundary conditions. A particle $\mu$ at site $i$ attempts hops forward
to site $i+1$
with rate $p_{\mu}$ and hops backward to site $i-1$ with rate $q_{\mu}$.
Due to exclusion,  hops  only succeed if the target site is empty.
Since particles cannot overtake the sequence of particles is 
preserved.
We  summarise
the allowed exchanges at a pair of sites $i, i+1$ as 
\begin{equation}
 \mu\ 0 \rightarrow 0\ \mu \;\;\; \mbox{with rate}\;\;\;p_{\mu}
\;\;\; ; \;\;\;
0\ \mu \rightarrow  \mu\ 0 \;\;\;\mbox{with rate}\;\;\;q_{\mu}
\end{equation}
In order to obtain the steady state of this model 
a matrix product \cite{DEHP} was used. The ansatz is to represent the weight
(unnormalised probability) of a configuration
by the trace of a product of matrices $D_{\mu},E$, one matrix for each site;
we write $D_{\mu}$ for site $i$ if particle $\mu$ is at site $i$ in the
configuration and $E$ if a hole is at site $i$. For example,
the weight $f_N( n_1,n_2,\ldots, n_M)$ of a configuration of $N$ sites
comprising 
 particle 1 followed by $n_1$ holes; particle 2 followed by $n_2$ holes
and so on, is written as
\begin{equation}
 f_N( n_1,n_2,\ldots, n_M)=\mbox{Tr} \left[
D_1\ E^{n_1}\ D_2\ E^{n_2}\ \ldots D_M\ E^{n_M} \right]
\label{matf}
\end{equation}
The weights are related to probabilities
$P_N( \{n_{\mu}\})$ via a normalisation $Z_{N,M}$ 
defined by $P_N( \{n_{\mu}\}) =f_N( \{n_{\mu}\})/Z_{N,M}$.
We see from (\ref{matf}) that in matrix form
\begin{equation}
Z_{N,M} =
 \sum_{n_1,n_2 \ldots n_M}
\delta_{  \sum_{\mu} n_\mu,\ (N-M)}
\mbox{Tr} \left[ \prod_{\nu=1}^{M} D_{\nu} E^{n_\nu} \right]
\label{norm}
\end{equation}
 (so that the total number of holes is $N-M$).
A sufficient condition for 
(\ref{matf}) to indeed give the steady state is
\begin{equation}
p_{\mu} D_{\mu} E - q_{\mu} E D_{\mu} = D_{\mu} 
\;\;\;\mbox{for}\;\;\; \mu= 1\ldots M
\label{alg}
\end{equation}
Details of the proof
 will be given elsewhere \cite{MEtbp}.
Due to the fact that the particles cannot overtake each other,
the  velocity $v$ (the
steady state average of the rate of  forward hops minus
the rate of backward hops) is the same for each particle
$\mu$. Taking $\mu = 1$
\begin{eqnarray}
v&=&  \sum_{n_1,n_2 \ldots n_M} \delta_{ \sum_{\mu} n_\mu, \ (N-M-1) }
\mbox{Tr} \left[\ \{ 
p_{1}D_{1}\ E - q_{1} E D_{1} \} 
E^{n_1}\prod_{\nu =2}^{ M} D_{\nu} E^{n_\nu}\  \right]    / Z_{N,M} \nonumber \\
&=&Z_{N-1,M}/Z_{N,M}
\label{vel}
\end{eqnarray}
where we have used (\ref{alg}) and (\ref{norm}).
The existence \cite{MEtbp} of 
a non-trivial representation of
(\ref{alg}) provides a check that the algebra (\ref{alg}) is consistent.
However, rather than using a representation to perform calculations
we can proceed directly using (\ref{alg}).
We begin
by using (\ref{alg}) for $\mu=M$ in (\ref{matf}) to write
$$
f_N( n_1,\ldots, n_M) - \frac{q_M}{p_M} f_N( n_1,\ldots,n_{M-1}+1, n_M-1)=
\frac{1}{p_M} f_{N-1}( n_1,\ldots,n_{M-1}, n_M-1).
$$
The procedure is continued  using, in sequence,  (\ref{alg})
 with $\mu=M-1,M-2 \ldots 1$ to obtain
$$
\left[1- \prod_{i=0}^{M-1}\frac{q_{M-i}}{p_{M-i}}
                          \right] f_N( n_1,\ldots, n_M)
 =                         \left[ \sum_{i=0}^{M-1} \frac{1}{p_{M-i}}
                           \prod_{j=M+1-i}^{M} \frac{q_j}{p_j}\right]
                           f_{N-1}( n_1,\ldots, n_M-1)
$$
The effect of this manipulation has been to commute a hole initially
 in front of the particle $M$ backwards one full turn around the ring.
The result is that the weight of a configuration of  size $N$
is expressed as a multiple
of the weight of a configuration of  size $N-1$
 with one hole less in front
of particle $M$.
Repeating the  commutation procedure
for a hole initially in front of 
particles $M-1,M-2,\ldots 1$ we deduce that,
to within a multiplicative constant, the  weights
are
\begin{equation}
f_N( n_1,\ldots, n_M) = \prod_{\mu=1}^{M} g_{\mu}^{n_{\mu}}\;\;\;
\mbox{where}\;\;\; g_{\mu} =\left[
                         \sum_{i=0}^{M-1} \frac{1}{p_{\mu-i}}
                           \prod_{j=\mu+1-i}^{\mu} \frac{q_j}{p_j}
                           \right]
\ \left[1-\prod_{k=1}^{M}\frac{q_{k}}{p_{k}}
                          \right]^{-1} \; .
\label{ss}
\end{equation}

An immediate consequence of the form (\ref{ss})
is a  mapping to an ideal Bose gas.
The $N-M$ holes of the exclusion process are viewed
as Bosons which may reside in $M$ states with energies $E_{\mu}$
 determined by
the hopping rates of the $M$ particles i.e.
$\exp(-\beta E_{\mu}) = g_{\mu} $.
A hole  in the $\mu^{\it th}$ Bose state
is in front of particle $\mu$ (and behind particle $\mu +1$).
The normalisation $Z_{N,M}$ is equivalent
to the canonical partition
function of a Bose gas. The 
velocity is a ratio
of partition functions of different system sizes (\ref{vel})
and corresponds  to a fugacity.

One sees from (\ref{ss}) that the energy levels are, in general,
complicated functions of the hopping rates. However,
a simplification occurs if the particles are restricted to hop forward
only i.e. $q_{\mu}=0 \ \forall \mu$.
In that case the algebra (\ref{alg}) reduces to 
$p_{\mu} D_{\mu} E  = D_{\mu} $ and it is easy to see
that
$g_{\mu} =1/p_{\mu}$. For simplicity this case will be discussed
in the following.

\noindent {\bf An illustrative example of Bose condensation: }
To illustrate the phenomenon of Bose condensation in this model
consider first the simplest case of an inhomogeneous system:
 particle 1 has hopping rates $p_1=p\ , \ q_1=0$
while the other $M-1$ particles have hopping rates 
$p_{\mu}=1\;\; q_{\mu}=0$ for $ \mu >1$. This situation is, in fact, a limit
of the problem of a single moving defect studied recently \cite{BerChine}.
The normalisation $Z_{N,M}$ is easy to calculate combinatorically
since $g_1 = p^{-1}$ and for $\mu >1$, $g_{\mu} =1$ thus
\begin{equation}
 Z_{N,M} =
 \sum_{n_1=0}^{N-M}
\left( \begin{array}{c} N-n_1-2 \\ M-2 \end{array} \right) p^{-n_1} \; ,
\label{z1slow}
\end{equation}
yielding an exact expression for the velocity through
(\ref{vel}).
In the thermodynamic limit 
\begin{equation}
N \rightarrow \infty \;\;\;\mbox{with}\;\;\; M =\rho N\; ,
\label{thermlim}
\end{equation}
where   the density $\rho$ is held fixed,
the sum (\ref{z1slow}) is dominated by 
$n_1 \sim {\cal O}(N)$ for $\rho < 1-p$ and $n_1 \sim
{\cal O}(1)$ for $\rho > 1-p$ and it can be shown that
\begin{eqnarray}
\mbox{for}\;\;\; \rho<1-p \;\;\; &Z_{N,M} \simeq  p^{-(N-M)} \ (1-p)^{-(M-1)}&
 \;\; \mbox{and}\;\; v \rightarrow p \\
\mbox{for}\;\;\; \rho>1-p \;\;\; &Z_{N,M} \simeq
\left( \begin{array}{c} N \\ M \end{array} \right)\rho^2p/(\rho-1+p)&
 \;\; \mbox{and}\;\; v \rightarrow 1-\rho 
\end{eqnarray}
Two phases are apparent: a low density phase $\rho<1-p$ where the velocity
is equal to $p$, hence determined by the slowest particle; 
a high density phase $\rho>1-p$ where the velocity is $1-\rho$, hence
determined by the density. The low density phase correpsonds 
to a Bose condensate since in the steady state $\langle n_1 \rangle/N$
(the fraction of the holes  in front of the slowest particle)
is non-zero. Equivalently, the density of holes at 
the site immediately in front of the slowest particle is 1.
In the high density phase the density of holes at this site is 
$(1-\rho)/p$. At the transition  ($\rho_c = 1-p$) one may think of the
tailback behind the slow particle reaching all the way round the ring.
Above $\rho_c =1-p$ the velocity decreases from p according to
$v = p - (\rho-\rho_c)^{\alpha}$ where $\alpha =1$.

\noindent {\bf The  Disordered Case: }
We now turn to the general problem where each particle's hopping
rate $p_{\mu}$  is a random variable.
To keep the analysis simple we assume particle 1 has the lowest hopping
 rate which we take to be $p_1=1$ while the other particles
have hopping rates drawn from a  distribution ${\cal P}(p)$.
 We
require that ${\cal P}(p)$ vanish at $p=1$ (so that particle 1 is indeed the
slowest) and as $p \rightarrow
\infty$, ${\cal P}(p)$ vanish fast enough that the distribution
 be normalisable.
Here a simple choice related to the gamma distribution
is considered
\begin{equation}
{\cal P}(p) =  (p-1)^{\gamma} \exp ( -(p-1) ) \left[ 
\Gamma(\gamma +1) \right]^{-1} 
\label{dist}
\end{equation}
where $\Gamma (\gamma)$ is the usual factorial function.
Other distributions may be studied \cite{MEtbp}.

To analyse the disordered case one wishes to obtain  properties
given by  a typical realisation of the disorder (here
the particle hopping rates),
for example
the typical velocity. Usually in the theory
of disordered systems one expects quantities such as $v$ to self-average
but quantities exponentially large in the system size, such as $Z_{N,M}$,
not to. That is, one expects as $N\rightarrow \infty$,
$v \rightarrow \overline{v}  \neq \overline{Z}_{N-1,M}/
\overline{Z}_{N,M}$ where the bar indicates
an average over the particle hopping rates. To circumvent this difficulty
 it is convenient to work with the
grand canonical ensemble \cite{Huang}
where in the thermodynamic limit,
we have (\ref{vel})
\begin{equation}
v=z \;\;\; \mbox{where} \;\;\; N-M = \sum_{\mu=1}^{M}
\frac{z g_{\mu}}{(1-z g_{\mu})}
\label{vgce}
\end{equation}
 In the thermodynamic limit
(\ref{thermlim})
the fraction of particles with hopping rates between
$p$ and $p+dp$ converges to ${\cal P}(p) dp$. Therefore (\ref{vgce}) 
may be replaced by
\begin{equation}
1-\rho = \rho I(z) +\frac{\langle n_1  \rangle}{N} \;\;\;\mbox{where}\;\;\;
I(z)=\int_{1}^{\infty} dp \ {\cal P}(p) \frac{z}{p-z}
\label{gce2}
\end{equation}
and $\langle n_1 \rangle = z/(1-z)$ denotes the steady state average
of $n_1$ (the number of holes in the Bose ground state).
Clearly, ${\cal P}(p)$ is analogous to the density of states
of the Bose gas.
The condition for Bose condensation \cite{Huang} translates
to
\begin{equation}
I(1) < (1-\rho)/\rho\; .
\label{cond}
\end{equation}
Recall from the theory of Bose condensation \cite{Huang} that
 since we must have $z\leq 1$, (\ref{cond}) implies 
$\langle n_1 \rangle /N$ is non-zero and $z=1$. If (\ref{cond}) does not
 hold, then in (\ref{gce2}),
 $\langle n_1 \rangle /N = 0$ and one solves for $z< 1$.
The integral $I(z)$, with ${\cal P}(p)$ given by
(\ref{dist}), may be expanded in powers of  $1-z$ via
a Mellin transform \cite{MEtbp} to obtain
\begin{eqnarray}
\mbox{for}& 0 < \gamma < 1  &
 I(z) =1/ \gamma
+ \Gamma(-\gamma)\ (1-z)^{\gamma} + \ldots\\
\mbox{for} & \gamma > 1 &
 I(z) = 1/ \gamma
- 1/(\gamma-1)\ (1-z) +\ldots
\end{eqnarray}
Hence, by (\ref{cond}) one has
$\rho_c =  \gamma /(1 +  \gamma)$
and  just above $\rho_c$, $v$ decreases as
\begin{eqnarray}
\mbox{for}& 0 < \gamma < 1 & \;\;\;
v \simeq 1 - \left[ - \Gamma(-\gamma) \rho_c^2
   \right]^{-1/\gamma}\ (\rho-\rho_c)^{1/\gamma} \\
\mbox{for}&  \gamma > 1 & \;\;\;
v \simeq 1 - (\gamma-1) \rho_c^{-2} \ (\rho-\rho_c) 
\end{eqnarray}
For  $\rho <\rho_c$  we have Bose condensation:
the fraction of holes in front of the 
slowest particle  is
 $\langle n_1 \rangle/(N-M) = 1- I(1) \rho/(1-\rho)$.
At the transition this fraction vanishes and above the transition
 the velocity decreases from
1 as $1-v \sim (\rho- \rho_c)^{\alpha}$,
where $\alpha = 1/\gamma$ for $\gamma <1$
and $\alpha =1 $ for $\gamma >1$. For $\gamma > 1$ 
the same $\alpha$ as for a single slow particle
is found.

\noindent {\bf Discussion:}
In this work it has been shown that the steady state velocity
in a disordered exclusion model undergoes a phase transition
exactly  equivalent to Bose condensation.
In the low density phase (Bose condensate) the
velocity is equal to the hopping rate of the
slowest particle. Above a critical density a high density
phase is entered and
the velocity decreases according to an
exponent which
has been calculated exactly
here for the totally asymmetric case and  a specific
 distribution of hopping rates
(\ref{dist}). 
The exponent depends 
on the distribution of slow hopping rates
which may be considered analogous to the density of states
of the Bose gas. Other distributions of hopping rates
yield a similar behaviour \cite{MEtbp}.

In the latter part of this letter the scope was 
restricted to a totally asymmetric exclusion 
process where only forward hops are allowed. This implied that
in the  Bose gas the energies 
were independently distributed, 
reminiscent of a  Random Energy Model of disordered systems \cite{REM}. 
It would be of interest to investigate more fully the
partially asymmetric case where backwards hops ($q_{\mu}\neq 0$)
are allowed.
As is seen from (\ref{ss}), in that
case the energies of states in the Bose
gas are correlated. 
In the present work the ground state, i.e. the slowest particle,
was taken to be non-degenerate. In the case of  high degeneracy
of the slowest particle, such as only two hopping rates of particle,
or of a distribution ${\cal P}(p)$ peaked at $p=1$
the phase transition  is not present \cite{MEtbp}.

In the introduction the model was discussed as a caricature
of traffic flow. More realistic but  related 
traffic models based on asymmetric
exclusion with  parallel updating  
have been studied \cite{Traffic}. Interestingly, bunching
of cars has been  noted in these models \cite{Naga} and it is
 important to determine whether this is related to the Bose
condensation analysed here.

It is a pleasure to thank 
the Weizmann Institute 
for warm hospitality  during the final writing of this work
and B. Derrida for helpful discussions.

%
%

\end{document}